
\input harvmac.tex
\input tables.tex
%
%
%
%
%
%
%
%
\def\nopage0{\pageno=0 \footline={\ifnum\pageno<1 {\hfil} \else
              {\hss\tenrm\folio\hss}   \fi}}
%
%

%
%
\def\sqr#1#2{{\vcenter{\hrule height.#2pt
      \hbox{\vrule width.#2pt height#1pt \kern#1pt
         \vrule width.#2pt}
      \hrule height.#2pt}}}

%
%
\def \d2dots{\mathinner{\mkern1mu\raise1pt\vbox{\kern7pt\hbox{.}}\mkern2mu
\raise4pt\hbox{.}\mkern2mu\raise7pt\hbox{.}\mkern1mu}}
%
%
\def\smdrct#1{{\vcenter{\hbox{\vrule width.4pt height#1pt}}\kern-1.5pt\times}}

%
%

%
%
\def\cA{{\cal A}}          \def\cB{{\cal B}}          
                    
\def\cG{{\cal G}}

                    \def\cR{{\cal R}}
                    \def\cU{{\cal U}}

%
%

\def\CC{\rlap {\raise 0.4ex \hbox{$\scriptscriptstyle |$}}
\hskip -0.1em C}
\def\FF{\hbox to 8.33887pt{\rm I\hskip-1.8pt F}}
\def\NN{\hbox to 9.3111pt{\rm I\hskip-1.8pt N}}
\def\PP{\hbox to 8.61664pt{\rm I\hskip-1.8pt P}}
\def\QQ{\rlap {\raise 0.4ex \hbox{$\scriptscriptstyle |$}}
{\hskip -0.1em Q}}
\def\RR{\hbox to 9.1722pt{\rm I\hskip-1.8pt R}}
\def\ZZ{\hbox to 8.2222pt{\rm Z\hskip-4pt \rm Z}} 
\def\demi{{1\over 2}}
%
%
\def\typeA{type ${\cal A}$}
\def\typeB{type ${\cal B}$}
\def\Spin{{\cal S}{\rm pin}\,}
\def\IndA{{\cal I}{\rm nd}_{\cal A}\,}
\def\IndB{{\cal I}{\rm nd}_{\cal B}\,}
\def\Ind{{\cal I}{\rm nd}\,}
\def\TB{{\cal B}\,}
\def\uq{\cU_q(sl(2))}
%
%
%
\lref \rAedirne { D. Arnaudon, {\sl Fusion rules and $\cR$-matrix
for representations of $SL(2)_q$ at roots of unity,}
to appear in
Proceedings of the V$^{\rm th}$ Reg. Conf. on {\sl Mathematical Physics},
Trakya University, Edirne, Turkey (1991), eds. S. Baran and M. Koca. }
\lref \rAcomo { D. Arnaudon, {\sl Fusion rules and $\cR$-matrix
for the composition of regular spins with semi-periodic representations of
$SL(2)_q$,} Phys. Lett. {\bf B268} (1991) 217.}
\lref \rAkingston { D. Arnaudon, {\sl Quantum chains and
fusion rules for representations of $SL(2)_q$ at roots of unity,}
to appear in Proceedings of NSERC-CAP Workshop on
{\sl Quantum Groups, Integrable Models and Statistical Systems},
Kingston, Canada, July 1992, eds. J. LeTourneux and L. Vinet.  }
\lref \rAGL { A. Alekseev, D. Gluschenkov and  A. Lyakhovskaya,
{\sl Regular representation of the quantum group $sl_q(2)$ ($q$ is a
root of unity),}
Univ. Paris-VI preprint PAR-LPTHE 92-24 June (1992).}
\lref \rBab { O. Babelon,
{\sl Representations of the Yang--Baxter algebrae associated to Toda
field theory,} Nucl. Phys. {\bf B230} [FS10] (1984) 241.}
\lref \rDK {  C. De Concini and V.G. Kac,
{\sl Representations of quantum groups at roots of 1,}
Progress in Math. {\bf 92} (1990) 471 (Birkh\"auser).}
\lref \rDeAk {  T. Deguchi and Y. Akutsu,
{\sl Colored vertex models, colored IRF models and invariants of
trivalent colored graphs,}
Tokyo preprint PRINT-92-0241 (1992).}
\lref \rDJMM { E. Date, M. Jimbo, K. Miki and  T. Miwa,
{\sl New $\cR$-matrices associated with cyclic representations of
$\ \cU_q(A_2^{(2)})$}, Kyoto preprint RIMS-706 and
{\sl Generalized chiral Potts models and minimal cyclic
representations of
$\ \cU_q \left(\widehat{gl}(n,\CC )\right) $,}
Commun. Math. Phys. {\bf 137}  (1991) 133.}
\lref \rGRSGS { C. G\'omez, M. Ruiz-Altaba and G. Sierra,
{\sl New $\cR$-matrix associated with finite dimensional representations
of $\ \cU_{q}(SL(2))$ at roots of unity,} Phys. Lett. {\bf B265}
(1991) 95;
C. G\'omez and G. Sierra,
{\sl A new solution  of the Star-Triangle equation based on $\cU _q(sl(2))$
at roots of unit,}
Nucl. Phys. {\bf B373} (1992) 761
and
{\sl New integrable deformations of higher spin Heisenberg-Ising chains,}
Phys. Lett. {\bf B285} (1992) 126.}
\lref \rKel { G. Keller,
{\sl Fusion rules of $\ \cU_{q}(SL(2,\CC))$, $q^{m}=1$,} Letters in
Math. Phys. {\bf 21} (1991) 273.}
\lref \rKer { T. Kerler,
{\sl Quantum groups, quantum categories and quantum field theory,} Ph.D.
Thesis Diss., ETH 9828 (1992). }
\lref \rLus { G. Lusztig,
{\sl Quantum groups at roots of $1$,} Geom. Ded. {\bf 35} (1990) 89.}
\lref \rPS { V. Pasquier and  H. Saleur, {\sl Common
structure between finite systems and conformal field theories through
quantum groups,} Nucl. Phys. {\bf B330}  (1990) 523.}
\lref \rRA {  P. Roche and  D. Arnaudon,
{\sl Irreducible representations of the quantum
analogue of $SU(2)$,}
Lett. Math. Phys. {\bf 17} (1989) 295.}
\lref \rRosA { M. Rosso,
{\sl Finite dimensional representations of the quantum analogue of the
enveloping algebra of a complex simple Lie algebra,}
Commun. Math. Phys. {\bf 117} (1988) 581.}
\lref \rRuiz { M. Ruiz-Altaba, {\sl New solutions to the Yang--Baxter
equation from
two-dimensional representations of $\ \cU_q(sl(2))$ at roots of unit,}
Phys. Lett. {\bf B279} (1992) 326.}
\lref \rSal { H. Saleur, {\sl Representations of $U_q sl(2)$ for q a
root of unity,}
in Proceedings of {\sl Nuclear Theory and Physics}, Les Houches,
March 1989. }
\lref \rSkl { E. K. Sklyanin, {\sl Some algebraic
structures connected with the
Yang--Baxter equation. Representations of quantum algebras,}
Funct. Anal. Appl. {\bf 17} (1983) 273. }
%
%
\line{\hfil                                                   CERN-TH.6730/92}
\Title{}{\vbox{\centerline{Composition of Kinetic Momenta:}
      \vskip2pt\centerline{The $\cU_q(sl(2))$ Case}} }

\centerline{Daniel Arnaudon\footnote{$^*$}
{arnaudon@surya11.cern.ch}}
\footnote{}{On leave from CNRS, France}
\medskip
\centerline{Theory Division}
\centerline{CERN}
\centerline{1211 Gen\`eve 23, Switzerland}

\bigskip
The tensor products of (restricted and unrestricted) finite
dimensional  irreducible
representations of $\uq$ are considered for $q$ a root of
unity. They are decomposed into
direct sums of irreducible and/or indecomposable
representations.

\vfill

\leftline{CERN-TH.6730/92}
\leftline{November 1992}
\leftline{hep-th/9212067}
\leftline{Published in Communications in Mathematical
Physics {\bf 159}}

\baselineskip=17pt

\Date{}

\newsec{Introduction}
When the parameter of deformation $q$ is not a root of unity, the
theory of representations of quantum algebras $\cU_q(\cG)$ (with $\cG$ a
semi-simple Lie algebra) is equivalent to the classical theory \rRosA.
In the following, we consider $\uq$, with $q$ a root of unity.
In this case, the dimension of the finite dimensional irreducible
representations (irreps) is bounded, and a new type of representations
occurs, depending on continuous parameters \refs{\rRA \rSkl \rBab  {--}
\rDK}. Moreover, finite dimensional representations are not always
direct sums of irreps: they can contain
indecomposable sub-representations. Some kinds of indecomposable
representations actually appear in the decomposition of tensor
products of  irreps.

Another peculiarity with $q$ a root of unity is that the fusion rules
are generally not commutative.
There exist however many sub-fusion-rings that
are commutative. The well-known one is the fusion ring generated by
the irreps of the finite dimensional quotient of $\uq$
\refs{\rPS \rKel {--} \rKer}.
Families of larger commutative fusion ring that
contain the latter will also be defined later.

\medskip

The following section is devoted to definitions, to the description of
the centre of $\uq$, and finally recalls the classification of the
irreps of $\uq$.
The  irreps of $\uq$ can be classified into two types:
\item {--}{
The first type,
called \typeA\ in the following, corresponds to the deformations of
representations that exist in the classical case $q=1$.  These
representations are also called restricted representations since they
are also representations of the finite dimensional quotient of $\uq$.
(This quotient consists in imposing classical values to the enlarged
centre of $\uq$.)
}
\item {--} {
The second type, denoted by $\cB$, contains finite dimensional irreducible
representations that have no finite dimensional classical analogue.
They are generically characterized by three continuous complex
parameters, which correspond to the values of the generators of the
enlarged centre, and they all have the same dimension. (This property
is a particularity of $\uq$. At higher ranks, several dimensions are
allowed for irreps. The dimension remains however bounded.)
}

\medskip

Section 3 is a review of the fusion rules for \typeA\
or restricted irreps \refs{\rPS \rKel {--} \rKer}.
The fusion ring generated by the
\typeA\ irreps also
contains a class of indecomposable representations of dimension
called $\IndA$ representations in the following.

\medskip

Section 4 deals with the composition of \typeA\
(restricted) with \typeB\ (unrestricted) irreps. These tensor products
generically lead to sums of \typeB\ irreps.
For non-generic parameters,
these fusion rules also lead to a new class of indecomposable
representations called $\IndB$ representations.

\medskip

The composition of
\typeB\ irreps is the subject of Section 5.
The tensor product  of two \typeB\ irreps is generically reducible
into \typeB\ irreps. However, it can also contain $\IndB$ representations
when the components of the tensor product do not have generic
parameters. For sub-sub-generic cases, the indecomposable
representations $\IndA$ reappear, together with, in even more
particular cases, another type of indecomposable representations
denoted by  $\IndA'$.

\medskip

The results presented in Sections 3, 4, 5 are summarized in {\it Tables
1,2,3}.

\medskip

In Section 6, we prove that the fusion ring generated by the
irreducible representations closes with the indecomposable
representations $\IndA$, $\IndA'$ and $\IndB$.

\medskip

The results of Section 5 are finally used as an example in Section 7
in the decomposition of the {\it regular representation} of $\uq$.

\newsec{Definitions, centre, and irreducible representations}
\subsec{Definitions}
The quantum algebra
$\uq$ is defined by the generators $k$, $k^{-1}$, $e$, $f$,
and the
relations
\eqn\eSL{
\eqalign{
& kk^{-1}=k^{-1}k=1 ,\cr
& [e,f]={k-k^{-1} \over q-q^{-1}}, \cr }
\qquad \qquad
\eqalign{
& kek^{-1}=q^2 e ,\cr
& kfk^{-1}=q^{-2}f .\cr}}

The
coproduct $\Delta $ is given by
\eqn\eDelta {
\eqalign {
& \Delta(k)=k\otimes k \cr
& \Delta(e)=e\otimes 1      + k\otimes e \cr
& \Delta(f)=f\otimes k^{-1} + 1\otimes f \;,\cr  }}
while the opposite coproduct $\Delta '$ is $\Delta '= P \Delta P$, where $P$
is the permutation map $P x\otimes y =y \otimes x$.
The result of the composition of two representations
$\rho_{1}$ and $\rho_{2}$ of $\uq$ is the representation
$\rho=(\rho_{1}\otimes \rho_{2})\circ \Delta$, whereas the composition in
the reverse order is equivalent to
$\rho '=(\rho_{1}\otimes \rho_{2})\circ\Delta'$.

\subsec{Centre of $\uq$}
The usual $q$-deformed quadratic Casimir
\eqn\eCas{C=fe+(q-q^{-1})^{-2}\left( qk+q ^{-1}k^{-1} \right)}
belongs to the centre of $\uq$. When $q$ is not a root of unity, $C$
generates this centre.

In the following, the parameter $q$ will be a root
of unity.  Let $m'$ be the smallest integer such that $q^{m'}=1$.
Let $m$ be
equal to $m'$ if $m'$ is odd, and to $m'/2$ otherwise.

Then the elements
$e^m$, $f^m$, and $k^{\pm m}$ of $\uq$ also belong to the centre \rRA.
Together with $C$, they actually  generate the centre of $\uq$,
and  these generators are related by a polynomial relation
\rDK.
We write here this relation as follows: let $P_m$ be the polynomial
in $X$, of degree $m$, ($P(X)=X^m+...$), such that
\eqn\ePm{P_m(X)=
{2\over (q-q^{-1})^{2m}} T_m\left( \demi (q-q^{-1})^{2} X \right)}
where $T_m$ is  the $m^{\rm th}$  Chebychev polynomial of
the first kind
\eqn\eCheb{T_m(X)= \cos(m \arccos X).}
Then the relation becomes
\eqn\ePmC{P_m(C)=e^m f^m + q^m {k^m +k^{-m} \over (q-q^{-1})^{2m}} \;.}

\subsec{Finite dimensional irreducible representations of $\uq$}
We now recall the classification \rRA\ of the irreducible representations of
$\uq$.
The new facts (with respect to the classical case or to the case
$q$ not being a root of
unity) are that the
dimensions of the finite dimensional irreps
are  bounded by $m$, and that the
irreps of dimension $m$ depend on three complex continuous parameters.
In the following, we will call \typeA\   irreps those that have a classical
analogue (restricted representations) and \typeB\  irreps the others.
We will mostly use a module notation.

We will denote by $x$, $y$, $z^{\pm 1}$, and $c$ the
values of $e^m$, $f^m$, $k^{\pm m}$, and $C$ on irreducible representations.

The
$q$-deformed classical irreps (\typeA) are labelled by their half-integer
spin $j$, which is
such that $1 \le 2j+1 \le m$, and by another discrete parameter
$\omega = \pm 1$ \rLus.
They are given  by the basis
$\{w_{0},...,w_{2j}\}$ and, in a notation of module,
\eqn\eSpinj{
\cases {
k w_{p}=\omega q^{2j-2p} w_{p} & for $0\le p \le 2j$ \cr
f w_{p}=w_{p+1} & for $0\le p \le 2j-1$ \cr
f w_{2j}=0 \cr
e w_{p}=\omega [p][2j-p+1] w_{p-1} & for $1\le p \le 2j$ \cr
e w_{0}=0 \cr
}}
where as usual
\eqn\eQN{[x]\equiv {q^{x}-q^{-x} \over q-q^{-1}}.}
We denote this representation by $\Spin(j,\omega)$.
On it, the central elements $e^m$, $f^m$, $k^{m}$,
and $C$ take the values $x=y=0$, $z=(\omega q^{2j})^m=\pm 1$, and
$c= \omega (q-q^{-1})^{-2} \left( q^{2j+1} + q^{-2j-1} \right) $
respectively.

Note that the representation $\Spin(j,\omega=-1)$  can be obtained
as the  tensor product of $\Spin(j,1)$
by the one-dimensional representation  $\Spin(j=0,\omega)$.

\medskip

A {\it \typeB\  irrep} is an irreducible representation that has no finite
dimensional analogue when $q$ is equal to one.
It has dimension $m$. It is characterized by three complex parameters
$x$, $y$, $z$ corresponding to the values of $e^m$, $f^m$, $k^m$, and by a
discrete choice among $m$ values $c_l$ for the quadratic Casimir $C$.
These values are just the roots of
\eqn\eRelation{P_m(c)-xy-q^m {z+z^{-1} \over (q-q^{-1})^{2m}}=0 \;.}
If we define $\zeta$ by
\eqn\eZeta{xy+q^m {z+z^{-1} \over (q-q^{-1})^{2m}} =
{\zeta ^m + \zeta ^{-m} \over (q-q^{-1})^{2m}}, }
then,
by virtue of the identity
\eqn\eGrad{\cos m\psi -\cos m \phi = 2^{m-1} \prod_{k=0}^{m-1}
\left(\cos \psi - \cos(\phi+2k\pi/m)\right), }
the $c_l$'s are given by
\eqn\eCl{c_l={\zeta q^{2l} + \zeta^{-1} q^{-2l} \over (q-q^{-1})^{2}}
\qquad l=0,...,m-1.}
Let $\lambda$ be an $m^{\rm th}$  root of $z$ and $c$ one of the $c_l$'s.
Then
the \typeB\  representation,  denoted in the
following by $\TB(x,y,z,c)$,
is given, in the basis $\{v_{0},...,v_{m-1}\}$, by
\eqn\eTypeB{
\cases {
k v_{p}=\lambda q^{-2p} v_{p} & for $0\le p \le m-1$ \cr
f v_{p}= v_{p+1} & for $0\le p \le m-2$ \cr
f v_{m-1}=y v_0 \cr
e v_{p}=\left( c - {1 \over (q-q^{-1})^{2}} \left( \lambda q^{-2p+1} +
\lambda^{-1}  q^{2p-1} \right) \right) v_{p-1}
      & for $1\le p \le m-1$  \cr
e v_{0}={1 \over y} \left( c - {1 \over (q-q^{-1})^{2}} \left( \lambda q +
\lambda^{-1}  q^{-1} \right) \right)  v_{m-1}. \cr
}}

\medskip

{\it Remark 1: }
in this basis, the generators $e$ and $f$ do not play symmetric roles.
The normalizations of the vectors are such that $f$ is extremely
simple in this basis. There exist of course more symmetric bases, and
bases where $e$ has a simple expression (related to the latter by a
simple change of normalization). The advantage of this basis
is that it can describe (irreducible)
representations with two highest-weight
vectors ($e$ vanishes on two vectors of the basis)
and a non-vanishing $y$.
For cases where $y$ vanishes but not $x$, another basis could be
preferable.
However, the limit  $y \rightarrow 0$ is well-defined if
$ c = {\lambda q + \lambda^{-1}  q^{-1} \over (q-q^{-1})^{2}} $ and
$e v_0 = \beta v_{m-1}$, $\beta \in \CC$.

\medskip

The representation \eTypeB\ is actually irreducible iff  one of
the four following conditions is satisfied:
\itemitem {a)} {$x \neq 0 $, }
\itemitem {b)} {$y \neq 0 $, }
\itemitem {c)} {$z \neq \pm 1 $, }
\itemitem {d)} {$c = { 2\omega \over (q-q^{-1})^2} \quad (\omega=\pm 1)$. }

\medskip

{\it Remark 2:}
Note that $\TB(0,0,\pm 1,
{2\omega \over (q-q^{-1})^2})=\Spin((m-1)/2,\omega)$
(fourth case) is actually of \typeA. This case will not be considered
as \typeB\  in the following. So a \typeB\  irrep will have
$(x,y,z)\neq(0,0,\pm 1)$.

\medskip

{\it Remark 3:}
The representations described by \eTypeB\ with $(x,y,z)=(0,0,\pm 1)$,
and one of the other possible
values  for $c$ ($\beta$ arbitrary, cf. {\it remark 1}) are
indecomposable. These representations, called $\IndA'$,
will appear in the last section
as indecomposable parts of some tensor products.

\medskip
For further use, we define the function $c(\zeta)$ by
\eqn\eczeta{c(\zeta) \equiv  {\zeta +\zeta^{-1} \over (q-q^{-1})^2}\;.}
\medskip

The representation \eTypeB\ will be called periodic if $xy \neq 0$.
In this case it is irreducible and has no highest-weight
and no lowest-weight vectors.
A semi-periodic representation is a representation for which one only of the
parameters $x$ and $y$ vanishes.  It is then also irreducible.
Following \rGRSGS,
a \typeB\  representation with $x=y=0$, $z\neq \pm 1$ will be called
nilpotent.

\newsec {Composition of \typeA\  representations}

This section will be a brief review of the results of Pasquier and Saleur
\rPS,  of Keller \rKel, and of Kerler \rKer.
The tensor product of two representations $\Spin(j_{1},\omega_1)$
and $\Spin(j_{2},\omega_2)$
decomposes into
irreducible representations of the same type and also, if
$2(j_{1}+j_{2})+1$ is greater than $m$, into some indecomposable spin
representations.

An indecomposable  spin representation $\IndA(j,\omega)$ has dimension $2m$.
It is
characterized by a half-integer $j$ such that $1\le 2j+1 <m$
and by $\omega =\pm 1$.   In a basis
$\{w_{0},...,w_{m-1}, x_{0},...,x_{m-1}\}$ the generators of
$\uq$ act as follows~:
\eqn\eInd{
\cases {
k w_{p}=\omega q^{-2j-2-2p} w_{p} \cr
f w_{p}=w_{p+1} & for $0\le p \le m-2$ \cr
f w_{m-1}=0 \cr
e w_{p}=\omega  [p][-2j-p-1] w_{p-1} & for $0\le p \le m-1$ \cr
k x_{p}=\omega q^{2j-2p} x_{p} \cr
f x_{p}=x_{p+1} & for $0\le p \le m-2$ \cr
f x_{m-1}=0 \cr
e x_{p}=f^{p+m-2j-2} w_{0}+ \omega [p][2j-p+1] x_{p-1}
    & for $0\le p \le m-1\;.$\cr }}
(In particular, $e x_{0}=w_{m-2j-2}$ and $e x_{2j+1}=w_{m-1}$,
and $e^{m}$, $f^{m}$ are $0$ on such a module.)

This indecomposable representation contains the sub-representation
$\Spin(j,\omega)$. It is a deformation of the sum of the classical
$\Spin(j)$ and $\Spin(m/2-j-1)$ representations.

The fusion rules are
\eqn\eAAi {
\eqalign{
\Spin(j_1,\omega_1)\otimes \Spin(j_2,\omega_2)=&
\left( \bigoplus_{j=|j_1-j_2|} ^{\min(j_1+j_2,m-j_1-j_2-2)}
\Spin(j,\omega_1\omega_2) \right) \cr
&\bigoplus
\left( \bigoplus_{j=m-j_1-j_2-1}^{(m-1)/2} \IndA(j,\omega_1\omega_2)\right)
\;,\cr }}
where the sums are limited to integer values of $j$ if $j_ 1+j_2$ is
integer, and to half-(odd)-integer values if $j_ 1+j_2$ is
half-(odd)-integer. In conformal field theories, the fusion rules \eAAi\ are
truncated to the first parenthesis, keeping only those representations
that have
a $q$-dimension different from 0.

The fusion rules for  \typeA\ irreps
are summarized in {\it Table 1}.

The fusion rules of \typeA\  representations close with
\eqn\eAAii{\eqalign{
&\Spin(j_1,\omega_1)\otimes\IndA(j_2,\omega_2)=
\bigoplus_{{\rm some\;} j,\omega} \IndA(j,\omega)  \cr
&\IndA(j_1,\omega_1)\otimes\IndA(j_2,\omega_2)=
\bigoplus_{{\rm some\;} j,\omega} \IndA(j,\omega)  \;.\cr
}}
The $\Spin$ and $\IndA$ representations thus build a closed fusion
ring.

\newsec {Fusion rules mixing \typeA\  and \typeB\  representations }

{\bf Proposition 1: }{\sl
The tensor product of
a \typeB\  representation $\TB(x,y,z,c)$ with the  spin $1/2$
representation $\Spin(1/2,1)$ is
completely reducible iff
$c \neq {\pm 2 \over (q-q^{-1})^2}$.
More  precisely, if
$c = c(\zeta) = {\zeta + \zeta^{-1} \over (q-q^{-1})^2}$,
\eqn\eABi{
\TB(x,y,z,c) \otimes \Spin(1/2,1) =
\TB(x,q^m y,q^m z,c(q \zeta) )
\oplus \TB(x,q^m y,q^m z,c(q^{-1}\zeta))
\;.}
If $c = c(\pm 1)= {\pm 2 \over (q-q^{-1})^2}$,
the tensor product is a \typeB\
indecomposable representation of dimension $2m$, denoted by
$\IndB(x,q^m y,q^m z,c'= c(\pm q) =\pm{ q^{}+q^{-1} \over (q-q^{-1})^{2}})$
and defined below. }

{\it Proof: }
First write $c = {\zeta + \zeta^{-1} \over (q-q^{-1})^2}$.
The matrix of the quadratic Casimir on a weight space of
the tensor product is
diagonalizable iff $\zeta \neq \pm 1$, and the eigenvalues are
$c(q\zeta) \neq c(q^{-1}\zeta)$.
Each eigenvector of $C$
generates a \typeB\  irrep $\TB(x,q^m y,q^m z,c)$
since $(x,q^m y,q^m z)\neq (0,0,\pm 1)$.

When $\zeta=\pm 1$, the eigenvalues
$c(q\zeta)$ and $c(q^{-1}\zeta)$
coincide and $C$ is not
diagonalizable.
It has only one  eigenvector (up to a normalization) on each weight
space, which generates a \typeB\ irrep $\TB(x,q^m y,q^m z,c(\pm q))$.
The quotient of the total representation by this sub-representation is
again equivalent to  $\TB(x,q^m y,q^m z,c(\pm q))$. The tensor product
is hence the $2m$ dimensional indecomposable representation
$\IndB\left(x,q^m y,q^m z,c'=c(\pm q)\right)$.

{\it Definition:} The \typeB\ indecomposable representation
$\IndB(x,y,z,c)$ is characterized as follows:
the central elements
$f^m$ and $k^m$  take the scalar values $(y,z)$, and
there is a basis $\{v^{(i)}_{0},...,v^{(i)}_{m-1}\}$, ($i=1,2$), in which
this representation is written
\eqn\eTypeBind{
\cases {
k v^{(1)}_{p}=\lambda q^{-2p} v^{(1)}_{p} & for $0\le p \le m-1$ \cr
f v^{(1)}_{p}= v^{(1)}_{p+1} & for $0\le p \le m-2$ \cr
f v^{(1)}_{m-1}=y v^{(1)}_0 \cr
e v^{(1)}_{p}= \left( c- {\left( \lambda q^{-2p+1} +
          \lambda^{-1}  q^{2p-1} \right) \over (q-q^{-1})^{2}} \right)
          v^{(1)}_{p-1}
          & for $1\le p \le m-1$  \cr
e v^{(1)}_{0}= y^{-1} \left( c - {
          \left( \lambda q + \lambda^{-1}  q^{-1} \right)
          \over  (q-q^{-1})^{2}}
          \right)  v^{(1)}_{m-1} \cr
k v^{(2)}_{p}=\lambda q^{-2p} v^{(2)}_{p} & for $0\le p \le m-1$ \cr
f v^{(2)}_{p}= v^{(2)}_{p+1} & for $0\le p \le m-2$ \cr
f v^{(2)}_{m-1}= y v^{(2)}_0 \cr
e v^{(2)}_{p}=\left( c - {\left( \lambda q^{-2p+1} +
          \lambda^{-1}  q^{2p-1}  \right) \over (q-q^{-1})^{2}}
          \right) v^{(2)}_{p-1} + v^{(1)}_{p-1}
          & for $1\le p \le m-1$  \cr
e v^{(2)}_{0}=y^{-1} \left( \left( c - {
          \left( \lambda q + \lambda^{-1}  q^{-1} \right)
          \over  (q-q^{-1})^{2}}
          \right) v^{(2)}_{m-1} + v^{(1)}_{m-1} \right) \cr
}}
with
$\lambda ^m =z$.
We call this representation a \typeB\  indecomposable representation,
because $(x,y,z) \neq (0,0,\pm 1)$. It does not belong to the fusion
ring generated by the  \typeA\ irreps.

The sub-representation
generated by the set of $v^{(1)}_p$, as well as the quotient
of the whole representation by this sub-representation
are equivalent to
$\TB(x,y,z,c)$.

If $c=c(\zeta)$ with $\zeta^{2m}=1$ and $\zeta\neq \pm 1$ (which will
always be satisfied in the cases we will consider),  the
central element $e^m$ is scalar with value $x$ on
$\IndB(x,y,z,c(\zeta))$.
Otherwise, we would have
$$e^m v_p^{(1)} =x v_p^{(1)}\;,  \qquad
  e^m v_p^{(2)} =x v_p^{(2)} + {m\over y} {\zeta^m-\zeta^{-m} \over
\zeta -\zeta^{-1}} v_p^{(1)}\;. $$
In the following, we restrict the definition of  $\IndB$
representations to those representations that have one of the
special values for $c$ (i.e. $\zeta^{2m}=1$).
The operators $e^m$,
$f^m$ and $k^m$ hence take  scalar values on $\IndB$ representations.
As we will see in the next section, the property that these operators
are scalar on a representation is preserved in the composition of
representations. The fusion ring generated by the irreducible
representations then contains only representations with diagonal  $e^m$,
$f^m$ and $k^m$.

\medskip

The case $x=0$ and $y\neq 0$ (semi-periodic representation $\otimes$ spin
$1/2$) is included here. The description of the case $x\neq 0$ and
$y=0$ is simply obtained by considering bases with simple action of
$e$ instead of $f$. The case $x=y=0$ (nilpotent representation
$\otimes$
spin $1/2$) is included in this proposition and it does not lead to
indecomposability since the parameter $z,c$  (related to the highest
weight $\lambda$ through $z=\lambda^m$ and $c=c(q\zeta)$) of the
\typeB\ nilpotent representation
has to be generic (see {\it remark 2}).

\medskip

Let us again consider $\TB(x,y,z,c)$ with
$c = c(\zeta)$ \eczeta.
As a consequence of the previous proposition, we have:

\medskip

{\bf Theorem 1: }{\sl
The tensor product of
the \typeB\  representation $\TB(x,y,z,c)$ with the  spin $j$
representation $\Spin(j,1)$ is
completely reducible
as long as all the values
$c_l=c(q^{2j -2l}\zeta)$
for $l=0,...,2j$ are
different. (Which is satisfied in particular if $\zeta^{2m}\neq 1$.)
Moreover,
\eqn\eABii{
\TB(x,y,z,c) \otimes \Spin(j,1) = \bigoplus _{l=0}^{2j}
\TB\left(x,q^{2jm} y,q^{2jm} z,c_l=c(q^{2j -2l}\zeta)\right)
\;.}

The tensor product is not completely reducible when some pairs of
$c_l=c(q^{2j -2l}\zeta)$
($l=0,...,2j$) coincide.
(Since $2j+1\le m$, the $2j+1$ values $c_l$ can be only doubly
degenerate.)
In this case, the decomposition is obtained from \eABii\
by simply replacing  each pair of irreps arising
with the same $c_l$ by
the indecomposable \typeB\
sub-representation $\IndB(x,q^{2jm} y,$ $q^{2jm} z,c_l)$ \eTypeBind.
}

{\it Proof:} The previous proposition with the coassociativity of
$\Delta$ is the basic tool. The representation $\TB(x,y,z,c)$ is
composed with $(\Spin(1/2,1))^{\otimes 2j}$, which contains
$\TB(x,y,z,c) \otimes \Spin(j,1) $.
We however still need to know the result of the composition of
$\IndB(x,y,z,c)$ with $\Spin(1/2,1)$, since  $\IndB(x,y,z,c)$ can
appear in intermediate stages.

Let $c={\zeta + \zeta^{-1} \over (q-q^{-1})^2}$.
We look at the matrix of
$\Delta(C)$ on a weight space of the tensor product
$$\IndB(x,y,z,c) \otimes \Spin(1/2,1).$$
This matrix is a $4\times 4$ matrix. It can be decomposed into two
$2\times 2$ non-diagonalizable blocks with eigenvalues $c(q\zeta)$ and
$c(q^{-1}\zeta)$
if $\zeta$ is different from $\pm q$ and $\pm q^{-1}$.
If $\zeta=\pm q^{\pm 1}$, it can be decomposed into one $2\times 2$
non-diagonalizable block with eigenvalue $c(\pm q^2)$ and two $1\times
1$ blocks containing $c(\pm 1)$.
So the tensor product of
$\IndB(x,y,z,c)$ with $\Spin(1/2,1)$ reduces to
\eqn\eABiii{\IndB(x,y,z,c)
\otimes \Spin(1/2,1)=
\IndB(x,q^m y,q^m z,c(q\zeta))
\oplus
\IndB(x,q^m y,q^m z,c(q^{-1}\zeta)) }
if $\zeta$ is different from $\pm q$ and $\pm q^{-1}$,
and
\eqn\eABiv{\IndB(x,y,z,c)
\otimes \Spin(1/2,1)=
\IndB(x,q^m y,q^m z,c(\pm q^2))
\oplus 2\TB(x,q^m y,q^m z,c(\pm 1))
}
if $\zeta=\pm q^{\pm 1}$. The factor $2$ means a multiplicity of $2$ of
the representation in the decomposition, i.e. $\CC^2\otimes ...$

\medskip

\medskip

{\bf Proposition 2: }{\sl
If $\zeta^{2m}\neq 1$
the tensor product of
the \typeB\  representation $\TB(x,y,z,c)$ with a \typeA\
indecomposable representation $\IndA(j,1)$ is
completely reducible and
\eqn\eABv{
\TB(x,y,z,c) \otimes \IndA(j,1) = \bigoplus _{l=0}^{m-1}
2\ \TB\left(x,q^{2jm} y,q^{2jm} z,{q^{2j -2l}\zeta + q^{-2j+2l}\zeta^{-1}
\over (q-q^{-1})^2}\right)
\;.}
If $\zeta^{2m}= 1$, we have
\eqn\eABvi{
\TB(x,y,z,c) \otimes \IndA(j,1) = \bigoplus _{l=0}^{m-1}
\IndB \left(x,q^{2jm} y,q^{2jm} z,c(q^{2j -2l}\zeta) \right)
\;.}
with the prescription that $\IndB \left(x,q^{2jm} y,q^{2jm} z,c(\pm
1)\right)$, if it appears, has to be replaced by
$2\ \TB\left(x,q^{2jm} y,q^{2jm} z,c(\pm 1)\right)$.
(Such a prescription is of much easier use than an exploration of all
the cases: the parity of $m'$, $2j$ and the value of $\zeta$ enter in
the game.)

}
{\it Proof:}
The proof follows from the fact that $\IndA(j,1)$
enters in the decomposition of tensor products of some ordinary spin
irreps, as explained in the previous section.
This result is then obtained as the previous theorem by further
composition with the $\Spin(1/2,1)$ representation and using the
coassociativity of $\Delta$.
(Note that the reducibility obtained for  $\zeta^{2m}\neq 1$
holds although each root of the characteristic polynomial
of the quadratic Casimir is
doubly degenerate, whereas in the case of non-complete reducibility we
do not get $4m$-dimensional indecomposable representations.)

\medskip

The same technique leads to the decompositions of the tensor products
$\IndB \otimes \Spin$ and $\IndB\otimes \IndA$. We can actually
replace $\TB$ by $\IndB$ in \eABii\  and \eABv,\eABvi, always using
the prescription given for \eABvi.  (The representations
$\IndB(.,.,.,c(\pm 1))$ never
appear in our fusion rules, which is a key point for the closure of the
fusion ring.)

\medskip

We have only considered $\omega=1$ in the \typeA\  representations
entering in the fusion rules. We complete the fusion rules of \typeA\
with \typeB\  representations by adding
\eqn\eABomega{\TB(x,y,z,c)\otimes\Spin(0,-1)=
\TB(x,(-1)^m y,(-1)^m z,-c)\,.}

\medskip

These fusion rules were already considered in \rAcomo, in the cases
involving generic semi-periodic representations. The sub-cases leading
to indecomposability were however not considered.

\medskip

The decomposition of tensor products of \typeB\ irreps with \typeA\
irreps is summarized in {\it Table 2}. The cases involving the $\IndB$
and $\IndA$ representations are also summarized.

\medskip

One could remark here that the ``logarithm'' of the parameter $\zeta$
used in the expression of $c$ extends the role of the spin to
the case of \typeB\ representations:
the value of $\zeta$ for $\Spin(j,1)$ is $q^{2j+1}$, whereas
the tensor product by the
spin $1/2$ representation changes  $\zeta$  to $q^{\pm 1}\zeta$.
This is however not so simple in the following.

\newsec{Fusion of \typeB\  irreducible representations}

This section has many subsections. A summary of its content, including
the subsection numbers, is given in {\it Table 3}.

Consider two irreps of \typeB: $\rho_{1}=\TB(x_1,y_1,z_1,c_1)$  and
$\rho_{2}=\TB(x_2,y_2,z_2,c_2)$.

Then the central elements $e^m$, $f^m$, $k^{m}$ are scalar on
the tensor product  $\rho=(\rho_{1}\otimes \rho_{2})\circ \Delta$ and  take
the values
\eqn\exyz{
\eqalign{x&=x_1+z_1 x_2 ,       \cr
         y&=y_1 z_2^{-1}+y_2,   \cr
         z&=z_1 z_2.            \cr }}
They are also
scalar on  $\rho'=(\rho_{1}\otimes \rho_{2})\circ \Delta'$  and  take
the values  $(x'=x_2+z_2 x_1,y'=y_2 z_1^{-1}+y_1,z'=z_1 z_2)$.

\medskip

In fact, since
$$\eqalign{
\Delta(e)^m &= e^m \otimes 1      + k^m \otimes e^m \cr
\Delta(f)^m &= f^m \otimes k^{-m} + 1   \otimes f^m \cr
\Delta(k)^m &= k^m \otimes k^m \;,\cr
}$$
the fact that the
operators $e^m$, $f^m$ and $k^m$ are scalar is preserved by
the tensor product operation. Hence, since they are scalar on irreps,
they remain diagonal on the whole fusion ring generated by the irreps.

\medskip

We also see from \exyz\
that $\rho$ and $\rho'$ can be equivalent only if their parameters
belong to the same algebraic curve \rDJMM:
\eqn\eCurve{
{x_1\over 1-z_1}={x_2\over 1-z_2}
\quad ,\qquad
{y_1\over 1-z_1^{-1}}={y_2\over 1-z_2^{-1}}\;,}
and that in this case $x=x'$, $y=y'$, $z=z'$ also satisfy these relations.
In other words, since the coproduct is not co-commutative, the fusion
rules of representations are not commutative. If the values of the
parameters are restricted to belong to the same algebraic curve, the
corresponding restricted fusion rules are commutative.

For physical purposes, this condition will probably always be required.
However,
for more generality, we now consider the composition of $\rho_1$ and
$\rho_2$ with $\Delta$, without imposing the condition \eCurve.

The set of tensor products that we consider in this paper can
be restricted in such a way that the
representations belong to a given subset defined
by
\eqn\eCurveii{x=\hbox{const}(1-z) \qquad \hbox{and/or} \qquad
y=\hbox{const}'(1-z^{-1}).}
This subset of representations is stable under fusion.
Restriction of the fusion rules to this subset defines a
sub-fusion-ring that is commutative (when both conditions are
imposed).  The sub-ring generated by the
\typeA\ irreps is contained in these commutative sub-rings.
(The question of the closure of the fusion rings will be considered at
the end.)

\bigskip

Each weight space of
$\TB(x_1,y_1,z_1,c_1)\ \otimes \ \TB(x_2,y_2,z_2,c_2)$
has dimension $m$. The weights are all the $m^{\rm th}$ roots of $z=z_1z_2$.

\medskip
The following lemma is the main tool for all the further decompositions:

\medskip
{\bf Lemma 1:} {\sl
On a weight space on the tensor product
$$\TB(x_1,y_1,z_1,c_1)\ \otimes \ \TB(x_2,y_2,z_2,c_2),$$
the characteristic polynomial of $\Delta(C)$ is equal to the
polynomial
\eqn\eLemma{P_m(X)-xy-q^m {z+z^{-1} \over (q-q^{-1})^{2m}} \;,}
where $x$, $y$ and $z$ are given by \exyz. }

{\it Proof:}
The matrix of
\eqn\eDeltaC{
\Delta(C)=e\otimes f +fk\otimes k^{-1}e + C\otimes k^{-1} +k\otimes C
  -{q+q^{-1} \over (q-q^{-1})^2} k\otimes k^{-1}}
on a weight space is am $m\times m$
tridiagonal matrix (with three full diagonals, including two terms in
the corners).
The characteristic polynomial of this matrix is
then of degree $m$, and it contains basically two types of terms:
\item{--} { The
first type consists of the product of the elements of the upper diagonal
(respectively lower diagonal) elements. These two terms do not involve the
indeterminate $X$. They correspond to the values of $(e\otimes f)^m$
and $(fk\otimes k^{-1}e)^m$, i.e. $x_1y_2$ and $x_2y_1$.
}
\item{--} { The terms that involve at least one diagonal element of
the matrix of $\Delta(C)-X\cdot 1\otimes 1$. These  consist in
fact of products of diagonal elements with pairs of symmetric
off-diagonal ones.
The diagonal elements, which are evaluations of the last three terms of
\eDeltaC, depend on $c_i$ and $z_i$ only ($i=1,2$). The products of
symmetric off-diagonal elements have the same property, since the
products $ef$ and $fe$ are involved in their evaluation, not $e$ and
$f$ individually. }

So, one part of the constant term of the
characteristic polynomial of $\Delta(C)$ is
$(-1)^{m+1}(x_1y_2+x_2y_1)$ whereas the remaining terms only depend on
$c_i$ and $z_i$. The values $c_i$ are related with the products
$x_iy_i$ through \eRelation, but it is clear that we can vary $x_i$
and $y_i$ in such a way that their products (and $c_i$) remain
constant. This proves that we can vary continuously the constant term
of the polynomial, keeping the other terms constant.
So this
polynomial has $m$ distinct roots for generic values of the parameters.
These roots are then the $m$ distinct values for $c$ allowed by \eRelation\
with the corresponding generic $(x,y,z)$.
The characteristic polynomial of $\Delta(C)$ is then equal to \eLemma\
for generic $(x,y,z)$.
Since the characteristic polynomial of $\Delta(C)$
on the tensor product
is continuous in the
parameters,
it is equal to the polynomial \eRelation\ for
all the values of the parameters of the
representations.

\medskip
We know that the roots of \eLemma\ are either simple,
or doubly degenerate. The tensor product will then always be
decomposable into a sum of representations of dimension $m$ or $2m$,
corresponding to the characteristic spaces of $C$ (each of them being
either irreducible, indecomposable or again decomposable).

\subsec {Generic case}

{\bf Theorem 2: }{\sl
Consider two \typeB\  irreps  $\TB(x_1,y_1,z_1,c_1)$ and
$\TB(x_2,y_2,z_2,c_2)$. Let $(x,y,z)$ be defined by \exyz,
and
$\zeta$ by \eZeta. If $\zeta$ is not a $2m$-root of 1 (generic case),
the tensor product
$\TB(x_1,y_1,z_1,c_1) \otimes \TB(x_2,y_2,z_2,c_2) $
is reducible and
\eqn\eBBi{
\TB(x_1,y_1,z_1,c_1)\otimes \TB(x_2,y_2,z_2,c_2)  =
\bigoplus _{l=0}^{m-1}
\TB\left(x,y,z,c_l=
c(\zeta q^{2l})={\zeta q^{2l} + \zeta^{-1} q^{-2l} \over
(q-q^{-1})^{2}} \right)
\;.}
}

{\it Proof: }
We first note that the assumption on $\zeta$ forbids $(x,y,z)=(0,0,\pm
1)$.
So the tensor product cannot contain \typeA\  irreps. The \typeB\
irreps involved in the decomposition will be related to eigenvalues of
the quadratic Casimir $C$ \eCas\ (by the way, today is St. Casimir's day!).
The previous Lemma identifies
the characteristic polynomial of $\Delta(C)$ with the polynomial
\eLemma,
which has only  simple roots if $\zeta^{2m}\neq 1$.
The eigenspaces of $C$ then have dimension $m$ and they correspond to
the \typeB\ irreps of \eBBi, which are the only $m$-dimensional
representations of $\uq$ with parameters $(x,y,z,c_l)$.

\medskip
{\it Remark 4:} This theorem shows that two tensor products of
\typeB\  representations leading to the same $(x,y,z)$ with
$\zeta^{2m}\neq 1$ are equivalent, since their decompositions are
identical.

\medskip
The generic case of composition of \typeB\  irreps is then reducibility
into \typeB\  irreps.
\medskip
{\it Remark 5: } in ref. \rDJMM,
the underlying quantum Lie algebra is the affine
$\cU_q\left(\widehat{SL}(N)\right)$.
Analogous tensor products are in this case
irreducible, in contrast with the present results. Remember that in our case
the dimension of irreps is bounded by $m$.

\subsec {Sub-generic cases}

We consider in this subsection the tensor product
\eqn\eTP{
\TB\left(x_1,y_1,z_1,c_1=c(\zeta_1)={\zeta_1  + \zeta_1^{-1}  \over
(q-q^{-1})^{2}}
\right)
\otimes
\TB\left(x_2,y_2,z_2,c_2=c(\zeta_2)={\zeta_2  + \zeta_2^{-1}  \over
(q-q^{-1})^{2}}
\right)}
leading to $(x,y,z)$ with $\zeta^{2m}=1$ \eZeta. (The generic case was
$\zeta^{2m} \neq 1$.)

\medskip
\noindent {\it {\the\secno.\the\subsecno.1:}} \qquad
$(x,y,z)\neq(0,0,\pm 1)$.

We first assume
$(x,y,z)\neq(0,0,\pm 1)$. All the values $c_l$ \eCl\ are now
doubly degenerate roots of the characteristic polynomial of
$\Delta(C)$ on any weight space, except
$c(\pm 1)={\pm 2 \over (q-q^{-1})^{2}}$,
which can occur at most once.

The characteristic spaces of $\Delta(C)$, which are sub-representations
of the tensor product, can have the following structure:

\item{--}{ If related to the eigenvalue  $c(\pm 1)$, it has dimension
$m$ and is equivalent to $\TB(x,y,z,c(\pm 1))$. In this case, there is
only one possibility.}
\item{--}{ If related to the eigenvalue $c_l\neq c(\pm 1)$, it has
dimension $2m$. The only possibilities in this case are
\itemitem {--} { the corresponding representation is equivalent to the
indecomposable representation $\IndB(x,y,z,c_l)$;}
\itemitem {--} { it is reducible into a sum of two representations
equivalent to $\TB(x,y,z,c_l)$.}
}

The study of some cases shows that the first possibility is generic,
whereas the second also exists for special values of the
parameters.

\medskip

{\it Conjecture: }
We conjecture that
the tensor product \eTP, in the case $\zeta ^{2m}=1$ \eZeta\  and
$(x,y,z)\neq (0,0,\pm 1)$ \exyz, is obtained from the decomposition
\eBBi\ by coupling the pairs of \typeB\ irreps $\TB(x,y,z,c_l)$ whose
values of $c_l$ coincide into \typeB\ indecomposable representations
$\IndB(x,y,z,c_l)$
\eTypeBind.
For special values of
the parameters, however, they can remain decoupled. A necessary
condition for this decoupling is that $\zeta_1$ and $\zeta_2$ are also
$2m$-roots of $1$.

\medskip
\noindent {\it {\the\secno.\the\subsecno.2:}} \qquad
$(x,y,z)=(0,0,\pm 1)$.

Consider now
$\TB(x_1,y_1,z_1,
c_1=c(\zeta_1) )
\otimes \TB(x_2,y_2,z_2,c_2=c(\zeta_2))$
leading to  $(x,y,z)=(0,0,\pm 1)$. We choose $z=+1$, the other case
being similar.
Thus $x_2=-z_1^{-1} x_1$, $y_2=- z_1 y_1$,
$z_2= z_1^{-1}$.
Applying Eq. \eZeta\  to each set of variables
$(x_1,y_1,z_1,c_1)$ and  $(x_2,y_2,z_2,c_2)$, we can fix
$\zeta_2=q^{2j_1} \zeta_1$
with $2j_1$ integer ($\leq m$).

\medskip
\noindent {\it {\the\secno.\the\subsecno.2.1:}} \qquad   $x_1 y_1 \neq 0$.

In this case, $\Delta(e)$ and $\Delta(f)$ have a rank equal to $m-1$
on each weight space of the tensor product.
In other words,
each weight space contains one and only one
highest-weight vector, and
also one and only one lowest-weight vector (up to normalization).

Each highest-weight or lowest-weight vector is an eigenvector of
$\Delta(C)$ (since it is an eigenvector of $\Delta(k)$).

{\bf Lemma 2:}

{\sl The $\Spin(j,\omega)$ irrep is a sub-representation of the tensor
product
$\TB(x_1,y_1,z_1,c_1) \otimes \TB(x_2,y_2,z_2,c_2)$
if and only if $\zeta_1 / \zeta_2$ or $\zeta_1 \zeta_2$
is a weight of $\Spin(j,\omega)$.}

{\it Proof:} Consider a vector of weight $\omega q^{2j}$ in the tensor
product, annihilated by $\Delta(e)$ (unique up to a normalization; its
computation is straightforward).
This vector is the only canditate as highest weight of
$\Spin(j,\omega)$. From the relations satisfied by the generators of
$\uq$, we know that the first power of $\Delta(f)$ that can annihilate this
vector is either $2j+1$ or $m$. In the first case (and in this case
only), the representation
$\Spin(j,\omega)$ is a sub-representation of the tensor product.
An explicit calculation proves that
the
condition for $\Delta(f)^{2j+1}$ to cancel our highest-weight vector
is then exactly
\eqn\eCondSpin{\prod_{l=-j,-j+1,...,j}
\left\{ \zeta_1+\zeta_1^{-1} -
\omega \left( \zeta_2 q^{2l} + \zeta_2^{-1} q^{-2l} \right) \right\}=0\;.}

In this subsection, we already fixed
$\zeta_2=q^{2j_1} \zeta_1$, but the Lemma 2 forces us
to consider again two cases:

\medskip
\noindent {\it {\the\secno.\the\subsecno.2.1.1:}} \qquad
$\zeta_1^{2m} \neq 1$.

Consider $(j,\omega)$ such that $q^{2jm}\omega^m=z$. In the case
$\zeta_1^{2m} \neq 1$, the preceding lemma proves that the tensor product
contains either $\Spin(j,\omega)$ or
$\Spin({m \over 2}-j-1,q^m \omega)$ (not both).

Each characteristic space of $\Delta(C)$ (of dimension $2m$)
then contains one, and only one, irreducible sub-representation, which
is of course of type $\Spin$ since $x=y=z^2-1=0$.
The only representation of $\uq$ of dimension $2m$, with weights of
multiplicity 2, with two highest-weight vectors, two lowest-weight
vectors and only one sub-irrep $\Spin(j,\omega)$ (or
$\Spin({m \over 2}-j-1,q^m \omega)$ respectively) is
$\IndA(j,\omega)$ (or
$\IndA({m \over 2}-j-1,q^m \omega)$ respectively).

We then have the following proposition:

\medskip
{\bf Proposition 3:} {\sl The tensor product
$\TB(x_1,y_1,z_1,c_1) \otimes \TB(x_2,y_2,z_2,c_2)$,
with
$$\eqalign{& x_1+z_1 x_2= y_1 z_2^{-1}+y_2=0 \;,\cr
           & z_1 z_2 = 1 \;, \cr
           & x_1 y_1\neq 0 \;, \cr
           & \zeta_2=q^{2j_1} \zeta_1\;, \cr
           & \zeta_1^{2m} \neq 1 \;, \cr } $$
is equivalent to the sum
\eqn\eBBii{
\bigoplus _{j=j_1,j_1+1,... \atop j\leq {m-1\over 2}}
\IndA(j,1)
\oplus
\bigoplus _{j={m\over 2}-j_1,{m\over 2}-j_1+1,... \atop j\leq
{m-1\over 2}} \IndA(j,q^m)  \;,  }
with by convention $\IndA((m-1)/2,\omega) \equiv \Spin((m-1)/2,\omega)$. }

Only \typeA\ representations appear in this decomposition. No
continuous
parameter survives in the result.

\medskip
\noindent {\it {\the\secno.\the\subsecno.2.1.2:}} \qquad
$\zeta_1^{2m} = 1$.

In this limit, some Clebsch-Gordan
coefficients related to the decomposition \eBBii\ diverge and the
equivalence does not hold.
The previous lemma shows that more
\typeA\  irreps ($\Spin(j,\omega)$) (than in \eBBii)
are sub-representations of the tensor product.
For some $(j,\omega)$, the irreps $\Spin(j,\omega)$ and
$\Spin({m \over 2}-j-1,q^m \omega)$ can both be sub-representations of our
tensor product. They appear in this case as sub-representations of the
same characteristic space of $\Delta(C)$.
In this case, the only possibility for the corresponding
characteristic space of $\Delta(C)$ is neither
$\IndA(j,\omega)$ nor $\IndA({m \over 2}-j-1,q^m \omega)$, which
contain only one sub-irrep, but the
direct sum
\eqn\eBBiii{
\IndA'(j,\omega,\beta)
\oplus
\IndA'\left( {m \over 2}-j-1,q^m \omega,\beta \right)  \,,
}
where $\IndA'(j,\omega,\beta)$ is  an $m$-dimensional indecomposable
representation\footnote{$^1$}
{These indecomposable representations were denoted by
$\IndB'$ in a previous version of this paper. We apologize for this
change of notation (note that there is no possible confusion)
motivated by the fact that $\IndA'$ representations
are representations of the finite dimensional quotient of $\uq$,
like \typeA\ irreps and $\IndA$ representations. They are actually
quotients of $\IndA$ representations.}
containing $\Spin(j,\omega)$ as sub-irrep, and
described by
\eTypeB\ with
$$\left( x=0,y=0,z=(\omega q^{2j})^m, c=c(\omega q^{2j+1}) \right),
\lambda=\omega q^{2j},$$
(respectively $\lambda= \omega q^{m-2j-2}$),
but $\beta \neq 0$ (see {\it Remarks 1} and {\it 3}).
These
representations  never appear in the fusion rules
of \typeA\  irreps for the following reason:
although they are not periodic  (they correspond
to  $x=y=0$), they share with periodic representations the fact that
$e^p$
and $f^{m-p}$ can have non-vanishing matrix elements between the same
vectors, in the basis of \eTypeB,  which diagonalizes $k$.
Moreover, unlike the previous case,
a continuous parameter ($\beta$ in {\it Remark 1})
remains in these representations, which depends on the parameters of
the initial representations. (After  all our constraints
are taken into account, two parameters remain, e.g. $y_1$ and $z_1$.)

The parameter $\beta$ in $\IndA'(j,\omega,\beta)$, which is the ratio
of the action of $e$ and $f^{m-1}$ on $e^{-1}\{ \ker f\}$, can be
considered as intrinsic and basis-independent.
The limit $\beta=0$ is well-defined and appears in the following. The
limit $\beta\rightarrow\infty$, which is the symmetric of
$\beta\rightarrow 0$ when the roles of $e$ and $f$ are exchanged, is
also well-defined but the representation has first to be written in
the basis where $e$, instead of $f$,  has a simple expression.

\medskip

Let $\zeta_1=q^{l_1}$, $\zeta_2=q^{l_2}$, $0\le l_i\le m-1$,
$2j_1=|l_2-l_1|$. Denote by  $2j_2$ either $l_1+l_2$ if
$l_1+l_2\leq m$, or $2m-l_1-l_2$ otherwise.

\medskip
{\bf Proposition 4:} {\sl With the data given above, the decomposition
is
\eqn\eBBiv{ \eqalign{
& \bigoplus _{j=\sup(j_1,j_2),\sup(j_1,j_2)+1,... \atop j\leq {m-1\over 2}}
\IndA(j,1)
\oplus
\bigoplus _{j={m\over 2}-\inf(j_1,j_2),{m\over 2}-
\inf(j_1,j_2)+1,... \atop j\leq {m-1\over 2}} \IndA(j,q^m)  \cr
& \oplus
\bigoplus _{j=\inf(j_1,j_2), \inf(j_1,j_2)+1,...}^{\sup(j_1,j_2)-1}
\left(
\IndA'(j,1,\beta) \oplus \IndA'\left( {m \over 2}-j-1,q^m,\beta \right)
\right)
\,, \cr
}}
for some $\beta$'s.}

\medskip
\noindent {\it {\the\secno.\the\subsecno.2.2:}} \qquad
$x_1 y_1=0$.

The results in this case are essentially the same as when
$x_1 y_1\neq 0$. However, they can be obtained through different
proofs, using  simpler expressions for the highest-weight and
lowest-weight
vectors of tensor products.

The representations involved in the tensor product \eTP\ are now
semi-periodic or nilpotent. In the case of a tensor product of semi-periodic
representations, we consider $x_1=x_2=0$, the case of lowest-weight
semi-periodic representations ($y_1=y_2=0$) being symmetric of the latter.
In this case,
their parameter $\zeta$ can be related to their highest-weight $\lambda$
through $\zeta_1=q\lambda_1$
and $\zeta_2=q^{-1} \lambda_2^{-1}$.

As for periodic representations, we have to distinguish two cases:

\medskip
\noindent {\it {\the\secno.\the\subsecno.2.2.1:}} \qquad
$\zeta_1^{2m}\neq 1$ (and hence $\zeta_2^{2m}\neq 1$)

In this case,
the ranks of $\Delta(e)$ and $\Delta(f)$ are still $m-1$ on each
weight space of the tensor product \eTP. So the number of highest- and
lowest-weight vectors on each characteristic space of $\Delta(C)$ is
the same as when $x_1y_1\neq 0$. Lemma 2 is still valid, and the
decomposition \eBBii\ still holds.

\medskip
\noindent {\it {\the\secno.\the\subsecno.2.2.2:}} \qquad
$\zeta_1^{2m}=1$ (and hence $\zeta_2^{2m}= 1$)

In this case, each
representation entering in the tensor product has two highest-weight
vectors, since the weights are $2m$-roots of $1$.
We consider only tensor products of irreps, so we must have no
lowest-weight vectors and hence $y_1 y_2\neq 0$.
(The representations are semi-periodic, not nilpotent.)

The rank of $\Delta(e)$  can now be $m-1$ or
$m-2$ on each weight space, depending on the weight, whereas the rank
of $\Delta(f)$ remains $m-1$ on each weight space.
If a highest-weight $q^{2j}$ is degenerate, we can check that the
weight $q^{-2j-2}$ also corresponds to two highest-weights.
Consequently, the characteristic space of $\Delta(C)$ that contains
them is equivalent to $\IndA'(j,1,0)\oplus \IndA'({m\over
2}-j-1,-1,0)$. For  the pairs of highest-weights $q^{2j}$ and
$q^{-2j-2}$  which are not degenerate, it is easy to see, from their
explicit expression, that one only is the highest-weight of a $\Spin$
sub-representation. This leads then
to the same decomposition  as for periodic
representations, i.e. formula \eBBiv\ with now vanishing $\beta$'s.

\medskip
The results of this section are summarized in {\it Table 3}.
\bigskip

Some of the fusion rules of \typeB\ irreps have already been
considered in the literature. In \refs{\rKel,\rKer,\rGRSGS,\rDeAk}, the
fusion of nilpotent representations was studied.
The generic case of fusion of semi-periodic irreps was considered in
\rGRSGS.
The fusion of generic periodic irreps for $q=i$ was described in
\rRuiz. Generic fusion rules were also presented in \rSal.
General results on fusion rules and $\cR$-matrices for $\uq$ were
given in \rAedirne,
and developed in \rAkingston.

\newsec{Fusion ring generated by all the irreps of $\uq$}

\medskip
{\bf Theorem 3:}

{\it The fusion ring generated by all the irreducible
representations of $\uq$ consists in
\item {--}{the irreducible representations of \typeA\ and $\cB$,}
\item {--}{the \typeA\ indecomposable representations $\IndA(j,\omega)$,}
\item {--}{the \typeB\ indecomposable representations
$\IndB(x,y,z,c(\zeta))$ (with $\zeta^{2m}=1$),}
\item {--}{the indecomposable representations of type
$\IndA'(j,\omega,\beta)$. }

This fusion ring contains sub-fusion-rings defined by imposing one or
both of the relations \eCurveii\ on the parameters $(x,y,z)$. When
both conditions are imposed, these
sub-rings are commutative.
}
\medskip
{\it Proof:} The previous results show that these four types of
representations are involved in the fusion ring. We still have to
prove that it closes without other types of representations.

The tensor products that have already been considered are
\item{--}{irrep $\otimes$ irrep}
\item{--}{$\IndA \otimes \Spin  \longrightarrow \IndA$ \qquad (Section 3)}
\item{--}{$\IndA \otimes \IndA  \longrightarrow \IndA$ \qquad (Section 3)}
\item{--}{$\TB   \otimes \IndA  \longrightarrow \TB $ or $\IndB$
                                \qquad (Section 4) }
\item{--}{$\IndB \otimes \Spin  \longrightarrow \IndB$ or $\TB(.,.,.,c(\pm 1))$
                                \qquad (Section 4) }
\item{--}{$\IndB \otimes \IndA  \longrightarrow \IndB$ or $\TB(.,.,.,c(\pm 1))$
                                \qquad (Section 4) }

(Reversed tensor products are similar, although not always equivalent.)

For the remaining tensor products, we will apply the following
procedure: we consider the indecomposable representations involved in
the tensor product as a term of the decomposition of a tensor product of
irreps. These irreps will always be chosen with the most generic
allowed parameters. The decomposition of the original tensor product
will then be a part of the decomposition of a tensor product of three
or four irreps, on which we will use the coassociativity of $\Delta$
(associativity of the fusion rules) and the previous results on the
composition of irreps. The first case will be treated in details, the
other being sketched.

\item{--}{$\TB \otimes \IndB$ with, on the result, $(x,y,z)$ and
$\zeta$, depending, as usual, on the original parameters.
Then
$$\eqalign{
\TB \otimes \IndB &\subset \TB \otimes (\TB_1 \otimes \TB_2)  \cr
                  &\subset (\TB \otimes \TB_1) \otimes \TB_2  \cr}$$
$\TB_1$ is considered as generic and the parameters of $\TB_2$ are
related to those of $\TB_1$ in order to contain $\IndB$ in their
fusion. Then $\TB\otimes \TB_1=\bigoplus \TB_3$, the irreps $\TB_3$
being as generic as $\TB_1$.
\itemitem{--}{
If $\zeta^{2m}\neq 1$, then $\TB_3\otimes\TB_2=\bigoplus\TB_4$, so
that
$$\TB\otimes \IndB\longrightarrow \bigoplus \TB\;.$$
}
\itemitem{--}{
If $\zeta^{2m}= 1$ and $(x,y,z)\neq (0,0,\pm 1)$, then
$$\TB_3\otimes\TB_2=\bigoplus\IndB {\hbox { and/or }} \TB_4,$$
so that
$$\TB\otimes \IndB\longrightarrow \bigoplus \IndB, \TB\;.$$
}
\itemitem{--}{
If $\zeta^{2m}= 1$ and $(x,y,z)= (0,0,\pm 1)$, then
$\TB_3\otimes\TB_2=\bigoplus\IndA $,
so that
$$\TB\otimes \IndB\longrightarrow \bigoplus \IndA\;.$$
}
}
\item{--}{$\TB(x,y,z,c(\zeta))\otimes \IndA'\subset \TB(x,y,z,c)\otimes\TB_1
\otimes \TB_2$.
The parameters $(x,y,z)$ of the result are those of
the \typeB\ irrep of the tensor product since the $\IndA'$
representations carries $(0,0,\pm 1)$.
\itemitem{--}{If $\zeta^{2m}\neq 1$, then $\TB\otimes \TB_1=\bigoplus \TB_3$
and $\TB_3\otimes\TB_2=\bigoplus \TB_4$, so that
$$\TB\otimes \IndA'\longrightarrow \bigoplus \TB\;.$$
}
\itemitem{--}{$\zeta^{2m}= 1$ (and $(x,y,z)\neq (0,0,\pm 1)$ otherwise
the first irrep is of \typeA), then $\TB\otimes \TB_1=\bigoplus \IndB,\TB_3$
and $(\IndB\hbox{ or }\TB_3)\otimes\TB_2=\bigoplus \IndB,\TB_4$, so that
$$\TB\otimes \IndA'\longrightarrow \bigoplus \IndB, \TB\;.$$
}
}
\item{--}{$\IndA'\otimes \Spin \subset \TB_1\otimes\TB_2\otimes\Spin$.
Since $\TB_2\otimes\Spin=\bigoplus\IndB,\TB$, and
$\TB_1\otimes (\IndB,\TB)=\bigoplus \IndA, \IndA'$, we have
$$\IndA'\otimes \Spin =\bigoplus \IndA, \IndA'.$$
}
\item{--}{The remaining cases, $\Ind \otimes \Ind$, with at most one
$\IndA$ in the tensor product, can be seen as included in
$\TB_1\otimes \TB_2 \otimes \Ind$, for which we use the previous cases.
}

The
conditions \eCurveii\ define sub-rings of the whole ring of
representations. Taking the intersection of the fusion ring generated
by irreps with these sub-rings provides interesting commutative
sub-fusion-rings.

\newsec{Decomposition of the regular representation of $\uq$}

Using \eBBii\ for nilpotent representations, we can
achieve the decomposition of the regular representation.

The
regular representation of $\uq$ is the finite dimensional
module defined by the left action  of $\uq$ on itself, with the
further relations $e^m=f^m=0$ and $k^{m'}=1$.

A natural basis is given by
$\{f^{r_1}e^{r_2}k^{r_3}\}$ with $r_1,r_2 \in \{0,...,m-1\}$ and
$r_3\in \{0,...,m'-1\}$. Using the basis
$\{v_{r_1,r_2,p}=\sum _{r_3=0}^{m'-1} q^{-r_3 p} f^{r_1}e^{r_2}k^{r_3}\}$
which diagonalizes the action of $k$, the regular representation was
decomposed in \rAedirne\ into the sum
\eqn\eRegi{
\bigoplus_{p=0}^{m'-1} \TB\left(0,0,\lambda^m,c(q \lambda)\right)
\otimes
\TB\left(0,0,\lambda^{-m},c(q^{p+1}\lambda^{-1})\right)
\;,}
which is then equivalent to
\eqn\eRegii{
\bigoplus_{j=0}^{(m-1)/2} (2j+1) \IndA(j,1)
{\oplus  \atop {\hbox {(if $m'$ is even)} }}
\left(
\bigoplus_{j=0}^{(m-1)/2} (2j+1) \IndA(j,-1)  \right)  \;.  }
We see that the multiplicity of each indecomposable representation is
equal to the dimension of its irreducible part.
Although \eRegi\ is valid for arbitrary $\lambda$, it is not a
surprise to find that the regular representation is of \typeA.
This result agrees with the decomposition obtained in \rAGL.

\bigskip
{\bf Acknowledgements}

\noindent
I thank A. Alekseev, D. Altschuler, O. Babelon, A. Coste,
R. Cuerno, A. LeClair, T. Miwa, R. Nepomechie,
V. Rittenberg, Y. Saint-Aubin, C. Viallet, L. Vinet and P. Zaugg
for  interesting
discussions.
The referee is also thanked for his fruitful remarks.

\listrefs

\null
\vfill
\begintable
$\Spin(j_1,\omega_1) \otimes \Spin(j_2,\omega_2)$   |
decomposes into                                               \crthick
$2(j_1+j_2)+1 \leq m$                               |
$\Spin (j,\omega_1\omega_2)$                                  \cr
$2(j_1+j_2)+1 >  m$                                 |
$\Spin (j,\omega_1\omega_2)$ and $\IndA(j,\omega_1\omega_2)$  \endtable

\bigskip
\centerline{\bf Table 1
\footnote{*}{\it Please insert table 1 at the end of Section 3}
:}
\centerline{Summary of the fusion rules for \typeA\ irreps}

\vfill

\begintable
$\TB_1$                                |  $\cA_2$                 |
decomposes into                              \crthick
$\TB$ irrep  with $\zeta_1^{2m}\neq 1$ |  $\Spin(j_2,\omega_2)$   |
$\TB$                                        \cr
$\TB$ irrep  with $\zeta_1^{2m}\neq 1$ |  $\IndA(j_2,\omega_2)$   |
$\TB$                                        \cr
$\TB$ irrep  with $\zeta_1^{2m}= 1$    |  $\Spin(j_2,\omega_2)$   |
$\IndB$ and/or $\TB(.,.,.,c(\pm 1))$         \cr
$\IndB$ rep  (with $\zeta_1^{2m}= 1$)  |  $\Spin(j_2,\omega_2)$   |
$\IndB$ and/or $\TB(.,.,.,c(\pm 1))$         \cr
$\TB$ irrep  with $\zeta_1^{2m}= 1$    |  $\IndA(j_2,\omega_2)$   |
$\IndB$ and/or $\TB(.,.,.,c(\pm 1))$         \cr
$\IndB$ rep  (with $\zeta_1^{2m}= 1$)  |  $\IndA(j_2,\omega_2)$   |
$\IndB$ and/or $\TB(.,.,.,c(\pm 1))$         \endtable

\bigskip
\centerline{\bf Table 2
\footnote{**}{\it Please insert table 2 at the end of Section 4}
:}
\centerline{Summary of the results of fusion of $\TB$ or
$\IndB$ representations with \typeA\ representations}

\vfill
\eject\null
\vfill

\def\adv{\hskip 1cm}
\def\hf{\hfill}

\begintable
section       \| $\TB_1$                     | $\TB_2$                     |
such that                      | decomposes into       \crthick
5.1       \hf \|                             |                             |
$\zeta^{2m}\neq 1$    \hf      | $\TB$                 \cr
5.2       \hf \|                             |                             |
$\zeta^{2m}= 1$       \hf      |                       \cr
5.2.1     \hf \|                             |                             |
\adv $(x,y,z)\neq (0,0,\pm 1)$ | $\IndB$, $\TB$        \cr
5.2.2     \hf \|                             |                             |
\adv $(x,y,z)=    (0,0,\pm 1)$ |                       \cr
5.2.2.1   \hf \| $x_1 y_1 \neq   0 $ \hf     | $x_2 y_2 \neq   0 $     \hf |
                               |                       \cr
5.2.2.1.1 \hf \| \adv  $\zeta_1^{2m}\neq 1 $ | \adv  $\zeta_2^{2m}\neq 1 $ |
                               | $\IndA$               \cr
5.2.2.1.2 \hf \| \adv  $\zeta_1^{2m}= 1 $    | \adv  $\zeta_2^{2m}= 1 $    |
                               | $\IndA$, $\IndA'$     \cr
5.2.2.2   \hf \| $x_1 y_1 = 0 $      \hf     | $x_2 y_2 = 0  $         \hf |
                               |                       \cr
5.2.2.2.1 \hf \| \adv  $\zeta_1^{2m}\neq 1 $ | \adv  $\zeta_2^{2m}\neq 1 $ |
                               | $\IndA$               \cr
5.2.2.2.2 \hf \| \adv  $\zeta_1^{2m}= 1 $    | \adv  $\zeta_2^{2m}= 1 $    |
                               | $\IndA$, $\IndA'$     \endtable

\bigskip
\centerline{\bf Table 3
\footnote{***}{\it Please insert table 3 at the end of Section 5}
:}
\centerline{Summary of the results of the fusion of \typeB\
irreps}

\vfill
\eject
\bye